# Learning optimal treatment strategies for intraoperative hypotension using deep reinforcement learning

Esra Adiyeke[a,b,*], Tianqi Liu[a,c,*], Venkata Sai Dheeraj Naganaboina [a,d,*], Han Li[a], Tyler J. Loftus[a,e], Yuanfang Ren[a,b], Benjamin Shickel[a,b], Matthew M. Ruppert[a,b] , Karandeep Singh[f], Ruogu Fang[a,g], Parisa Rashidi[a,g], Azra Bihorac[a,b,#], Tezcan Ozrazgat-Baslanti [a,b,#]

* These authors have contributed equally as first authors

# These authors have contributed equally as senior authors

[a] Intelligent Clinical Care Center (IC3), University of Florida, Gainesville, FL.

[b] Department of Medicine, Division of Nephrology, Hypertension, and Renal Transplantation, University of Florida, Gainesville, FL.

[c] Department of Electrical and Computer Engineering, University of Florida, Gainesville, FL.

[d] Department of Computer Science, University of Florida, Gainesville, FL.

[e] Department of Surgery, University of Florida, Gainesville, FL.

[f] Department of Medicine, University of California San Diego, San Diego, CA

[g] Department of Biomedical Engineering, University of Florida, Gainesville, FL.

**Corresponding author:** Azra Bihorac MD MS, Department of Medicine, Intelligent Clinical Care Center (IC3), Division of Nephrology, Hypertension, and Renal Transplantation, PO Box 100224, Gainesville, FL 32610-0224. Telephone: (352) 294-8580; Fax: (352) 392-5465; Email: abihorac@ufl.edu

# ABSTRACT

**Importance:** Traditional methods of surgical decision making heavily rely on human experience and prompt actions, which are variable. A data-driven system that generates treatment recommendations based on patient states can be a substantial asset in perioperative decision-making, as in cases of intraoperative hypotension, for which suboptimal management is associated with acute kidney injury (AKI), a common and morbid postoperative complication.

**Objective:** To develop a Reinforcement Learning (RL) model to recommend optimum dose of intravenous (IV) fluid and vasopressors during surgery to avoid intraoperative hypotension and postoperative AKI.

**Design, setting, participants:** We retrospectively analyzed 50,021 surgeries from 42,547 adult patients who underwent major surgery at a quaternary care hospital between June 2014 and September 2020. Of these, 34,186 surgeries were used for model training and internal validation while 15,835 surgeries were reserved for testing. We developed an RL model based on Deep Q-Networks to provide optimal treatment suggestions.

**Exposures:** Demographic and baseline clinical characteristics, intraoperative physiologic time series, and total dose of IV fluid and vasopressors were extracted every 15-minutes during the surgery.

**Main outcomes:** In the RL model, intraoperative hypotension (MAP<65 mmHg) and AKI in the first three days following the surgery were considered.

**Results:** The developed model replicated 69% of physician's decisions for the dosage of vasopressors and proposed higher or lower dosage of vasopressors than received in 10% and 21% of the treatments, respectively. In terms of intravenous fluids, the model's recommendations were within 0.05 ml/kg/15 min of the actual dose in 41% of the cases, with higher or lower doses recommended for 27% and 32% of the treatments, respectively. The RL policy resulted in a higher estimated policy value compared to the physicians' actual treatments, as well as random policies and zero-drug policies. The prevalence of AKI was lowest in the patients who received medication dosages that aligned with our agent model's decisions.

**Conclusions and Relevance:** Our findings suggest that implementation of the model's policy has the potential to reduce postoperative AKI and improve other outcomes driven by intraoperative hypotension.

**Introduction**

Postoperative acute kidney injury (AKI) affects almost 20% of the patients undergoing surgery and poses substantial risk for short- and long-term organ dysfunction and mortality.[1-7] Management of intravascular volume and vasomotor tone plays a substantial role in determining the risk for postoperative AKI.[6,8,9] Accordingly, prior research has shown associations between intraoperative hypotension and AKI.[10,11] Despite this importance, there are few consensus guidelines regarding best practices for predicting and treating intraoperative hypotension.[12] The Acute Disease Quality Initiative and Perioperative Quality Initiative recommended maintaining mean arterial pressure (MAP) > 65 mmHg based on moderate quality evidence, and recommended goal-directed blood pressure optimization based on stronger evidence.[5,13,14] Two randomized trials have demonstrated that goal-directed therapy reduces the odds of postoperative AKI and another demonstrated that machine learning-derived predictions of impending hypotension prompt anesthesiologists to act earlier, differently, and more frequently in managing intravascular volume and vasomotor tone, resulting in less time-weighted hypotension.[13,15]

Reinforcement learning (RL) is an artificial intelligence subfield that identifies the sequence of actions yielding the greatest probability of achieving a goal, such as normal postoperative renal function.[16] Previous works have applied RL algorithms to situations requiring real-time decisions accounting for a patient's continuously changing states[17-19], including sepsis,[20-23] sedation[24], pain management[25], diabetic glycemic management[26] and delirium[27]. Reinforcement learning has been particularly studied in sepsis management, where RL agents have been refined to account for model uncertainty,[28] multidimensional drivers of sepsis mortality,[29] and physician input to optimize fluid and vasopressor dosages.[30] In hypotension management, Futoma et al.[31] developed an RL approach to suggest multiple equivalent actions in an ICU setting. Zhang et al.[28] further refined RL optimization of ICU hypotension treatment by limiting an agent to only suggest actions at critical decision-making points, rather than at time intervals throughout the length of an ICU stay. However, there are currently no RL agents to minimize hypotension-related complications in the perioperative setting.

Our objective is to develop and validate a deep Q networks-based RL model that offers optimal dosing strategies for intravenous (IV) fluid and vasopressor administration for every 15-minute intervals during a major surgery to maintain optimal hemodynamic physiology and avoid postoperative AKI.

**Methods**

*Data Source*

The University of Florida Integrated Data Repository was the honest broker to assemble a single center longitudinal perioperative cohort for all patients admitted to the University of Florida Health (UFH) for patients of age 18 years or older during admission, following any type of major operative procedure between June 1st, 2014 through September 20th, 2020 by integrating electronic health records data from preoperative, intraoperative, and postoperative phases of care with other clinical, administrative, and public databases as previously described.[32] We excluded patients with end-stage kidney disease or underwent cardiac surgery, who died within 24 hours of surgery, hospital stay less than 24 hours, or had insufficient data. The final cohort consisted of 50,021 inpatient encounters who underwent major surgery (Supplemental Fig. 1). We chronologically split the cohort into development (admissions between June 1st 2014 and 29th November 2018, 70% of the entire cohort) and test (admissions between 30th November 2018 and 20th September 2020, 30% of the entire cohort) sets. We trained the model using the development cohort, allocating 30% for internal validation and parameter tuning, and reported model performance on the test set. If patients underwent multiple surgeries during an admission, only the first surgery was considered in the analyses. The University of Florida Institutional Review Board and Privacy Office approved this study with waiver of informed consent (IRB#201600223, IRB#201600262).

*Assessment of Kidney Function*

In determining AKI and chronic kidney disease (CKD), we considered Kidney Disease: Improving Global Outcomes (KDIGO) serum creatinine criteria.[33] We considered preadmission and admission serum creatinine records in determining baseline creatinine. We computed creatinine by back-calculation using the 2021 CKD-EPI refit without race equation assuming a baseline estimated glomerular filtration rate (eGFR) of 75 ml/min/per 1.73 m$^2$ in cases with no preadmission creatinine was available, in non-CKD patients.[34] We refer to a study by Ozrazgat-Baslanti et al. for all necessary details regarding the data elements, assumptions and algorithms of the computable phenotyping pipeline we developed and used.[35]

*Predictor Features*

The model development used 16 features, 4 static baseline features (age, sex, Charlson comorbidities index, body mass index), 9 intraoperative physiologic time series data (heart rate, systolic blood pressure, mean arterial pressure, body temperature, respiratory rate, minimum alveolar concentration, $SpO_2/FiO_2$ ratio, peak inspiratory pressure, and end-tidal carbon dioxide), 1 preoperative feature, which is the average mean arterial blood pressure of past 48 hours, and 2 additional features as cumulative IV fluid and cumulative pressor administered during surgery. We derived preoperative comorbidities from International Classification of Diseases (ICD) codes to calculate Charlson comorbidity indices.[36] We computed the mean values of vital signs from intraoperative time series data by resampling into 15-minute intervals. List of all input features and their statistical characteristics were given in Supplemental Tables 1 and 2.

*Reinforcement Learning Model*

The proposed RL algorithm is conceptualized as a dynamic model that utilizes preoperative and intraoperative features to recommend actions during surgery. We modeled for action prediction to prevent short term outcome of intraoperative hypotension (MAP<65 mmHg) and long-term outcome of AKI in the first three days following the surgery.[10]

The resulting actions (agent policy) are assessed compared to the actions taken by the physicians based on their experience (Fig. 1). This workflow simulates clinical tasks faced by physicians involved in perioperative care where patients' preoperative information is subsequently enriched by the influx of new data from the operating room. The final output produces the suggested actions in a dynamic manner, updating the suggestions every 15 minutes.

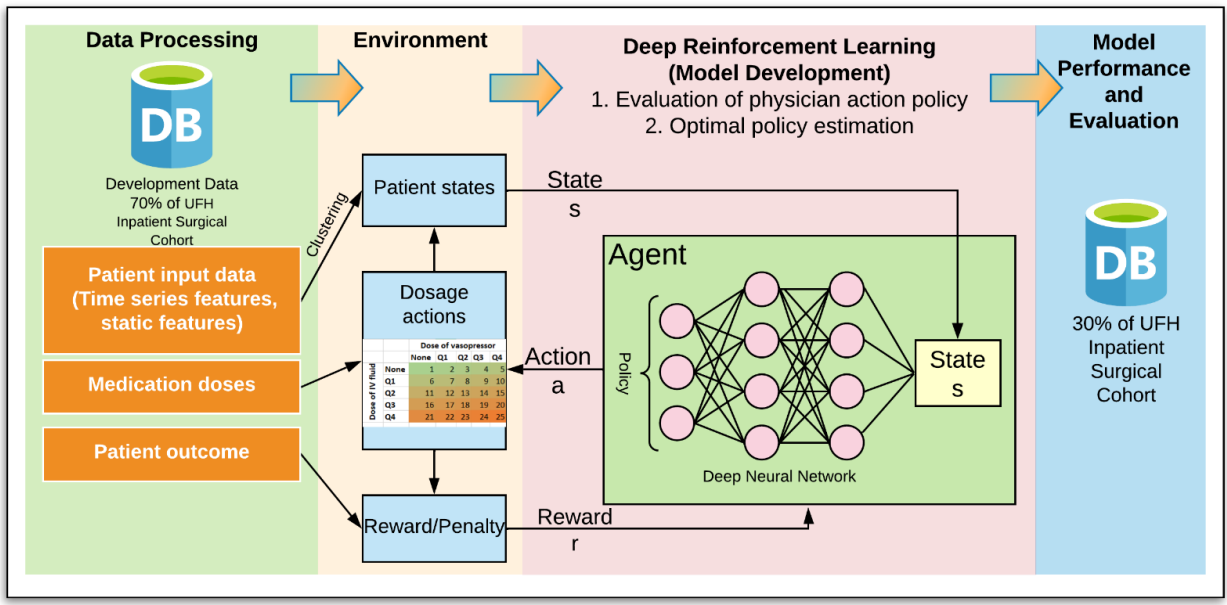

**Figure 1. Data flow and design of reinforcement learning approach using deep learning framework.** Data from our adult surgical cohort included static and times series features. We defined the action space with a total of 25 discrete actions, combining 5 different levels of IV fluid and vasopressor dosages. The model environment consists of EHR information, reward function, and associated actions. Environment supplies the observable states to the agent. Agent analyzes the state using a Deep Neural Network trained to produce the best action given optimum policy. The suggested action is sent to the environment, and the scored reward due to the action is sent back to the agent for training the policy. Model was trained and validated on 70% of the data and tested on 30% of the data, respectively.

The algorithm consists of two main layers, data processing and modeling, and each contains a data transformer core and a data analytics core.[32] Briefly, the workflow uses a data transformer to integrate data from multiple surgery rooms, and prepares the data for analysis through preprocessing and feature transformation (Supplemental Fig. 2). We used Dueling

Double Deep Q Networks (D3QN) as architecture for the RL model to test and analyze the optimal behavior. We discretized the state space by clustering the clinical variables which were resampled at 15-minute intervals as described in *Predictor Features*. Using the K-means++ clustering algorithm, we grouped patients to capture high-level similarities and then computed transition probabilities by considering the transition counts. We determined the number of clusters using silhouette analysis and selected 200 as the optimal value. We developed 50 distinct RL models for 50 different clustering outcomes obtained by different random initializations to account for the variability in policy values. We considered SHapley Additive exPlanations (SHAP) to evaluate the features' influences in RL model.[37] We used an enhanced version of the Deep Learning Important FeaTures (DeepLIFT) algorithm to approximate the SHAP values.[38] DeepLIFT can decompose the prediction of a neural network on a specific input by backpropagating the contributions of all neurons in the network to every feature of the input. We set the model running for 300 epochs, the batch size 256 experiences, and a learning rate of 1e-3.

*Action Space and Reward Function*

We converted the vasopressors using their norepinephrine equivalent in mcg/kg/min (Supplemental Table 3).[41] We discretized the action space into 5 groups using cutoff values derived from the empirical distribution of historical actions to ensure sufficient representation for each action category for IV fluids and vasopressors separately, one group corresponding to dosage of 0 (Supplemental Tables 4 and 5). We defined the action space with a total of 25 discrete actions, combining 5 different levels of IV fluid and 5 different levels of vasopressor dosages. We developed a reward function as a combination of long- and short-term rewards with two major parts to consider. In modeling the postoperative outcomes, we assigned the long-term reward (or penalty) utilizing a dedicated reward of +15 for AKI occurs within 3 days following the surgery, and -15 otherwise. The second part of the reward function adds a penalty of -1.75 if the state is hypotensive; that is if MAP is less than 65 mmHg or 20% lower than baseline MAP. We calculated the baseline MAP by calculating the median non-invasive MAP within the 48 hours prior to the first surgery start date-time that was between 60 and 110 mmHg.

*Performance Evaluation*

In RL, model performance is typically evaluated through interaction with a simulated environment using metrics like cumulative reward. However, due to lack of such environment in the current scenario, we assessed the model's policy quality through action prediction accuracy and off-policy evaluation techniques. To study that, we adopted visual analysis and weighted importance sampling (WIS) as quantitative method.[20,39,40] We adopted an off-policy evaluation to quantitatively assess our trained AI agent ($\pi_e$) based on physicians' policies. WIS takes the idea to re-weight the rewards in the historical data (physician's policy $\pi_b$) by the importance sampling ratio between $\pi_e$ and $\pi_b$. For visual analysis, policy scores of the model were compared to the physician policy relating the scores to outcome prevalence, along with distribution of actions in action space. We also evaluated the impact of difference in actions on the long-term outcome prevalence. To address the instability of evaluation performance in long-term cases, we propose a modified version of WIS for long-term applications. To clarify, we used logarithmic transformation combined with Softmax to constrain the range of the cumulative importance ratio, improving stability in long-term experiments (Supplemental Methods).

**Results**

*Patient Baseline Characteristics and Outcomes*

Among the 28,586 patients with 34,186 major surgeries in the development cohort, the mean age was 57 (standard deviation [SD], 17), 17,031 (50%) were female, 4,776 (14%) were African-American, 1,530 (4%) were Hispanic, and 15,578 (46%) had Medicare type of insurance (Table 1, Supplemental Table 2). The test cohort had 15,835 major surgeries from 13,961 patients. In this cohort, mean age was 59 (SD, 17), 7909 (50%) were female, 2,233 (14%) were African-American, 756 (5%) were Hispanic, and 7,672 (48%) had Medicare type of insurance.

The most common types of surgery, in descending order of frequency, were orthopedic, neurosurgical, and vascular procedures in both cohorts. The prevalence of postoperative complications in the development cohort was 11% for AKI within the first three days following surgery, 2% for 30 day-mortality and 4% for 90 day-mortality. In test cohort, percentage was 12% for postoperative AKI, 2% for 30 day-mortality and 3% for 90 day-mortality.

**Table 1. Clinical characteristics and outcomes of the patients.**

| Features | Development Cohort | Test Cohort |
|---|---|---|
| Number of encounters, n | 34,186 | 15,835 |
| **Demographic information** | | |
| Age, years, mean (SD) | 57 (17) | 59 (17) |
| Female, n (%) | 17,155 (50) | 7,926 (50) |
| African American, n (%) | 4,776 (14) | 2,233 (14) |
| Body Mass Index, median (IQR) | 28 (24, 34) | 28 (24, 34) |
| Emergency admission, n (%) | 12,243 (36) | 5,763 (36) |
| **Baseline clinical information** | | |
| Three most common types of surgery, n (%) | | |
|   Orthopedic surgery | 10,109 (30) | 4,341 (27) |
|   Neurosurgery | 3,744 (11) | 2,546 (16) |
|   Vascular surgery | 3,420 (10) | 1,704 (11) |
| Charlson comorbidity index, median (IQR) | 4 (2, 6) | 4 (2, 6) |
| Reference estimated glomerular filtration rate, median (IQR) | 95.27 (81.86, 109.73) | 93.44 (80.53, 107.71) |
| Baseline mean arterial pressure mmHg, median (IQR)[a] | 86 (78, 94) | 87 (79, 95) |
| **Intraoperative vitals, median (IQR)** | | |
|   Systolic blood pressure, mmHg | 114 (102, 130) | 116 (104, 132) |
|   Mean arterial pressure, mmHg | 79 (70, 90) | 82.0 (72, 93) |
|   Heart rate, bpm | 75.0 (65.50, 86.50) | 75.0 (66.0, 86.50) |
|   Oxygen saturation (SpO2), % | 99.10 (97.50, 10.00) | 99.00 (97.20, 10.00) |
|   Fraction of inspired oxygen (FiO2), % | 40 (40, 40) | 40.0 (40, 40) |
|   End-tidal carbon dioxide (EtCO2), mmHg | 34 (32, 37) | 35 (33, 38) |
|   Respiration rate, breaths/minute | 10 (8, 12) | 12 (10, 14) |
|   Peak inspiratory pressure, mmHg | 18.0 (14, 23) | 18.0 (14, 22) |
|   Minimum alveolar concentration (MAC) | 0.62 (0.44, 0.81) | 0.56 (0.31, 0.77) |
|   Core temperature, degrees Celsius | 36.83 (36.28, 37.33) | 36.94 (36.33, 37.44) |
| **Intraoperative medications, median (IQR)[b]** | | |
|   Vasopressor total dose per 15 min (mcg/kg) | 0 (0, 0.04) | 0 (0, 0.08) |
| Intravenous fluids total dose per 15 min (ml/kg) | 0.12 (0.03, 0.39) | 0.13 (0.03, 0.37) |
| **Intraoperative hypotension, n (%)** | 24,144 (70) | 9,677 (61) |
| **Postoperative AKI, n (%)** | | |
| AKI within 3 days after surgery | 3,870 (11) | 1,856 (12) |
| AKI within 7 days after surgery | 4,637 (14) | 2,242 (14) |
| AKI during hospitalization | 5,649 (17) | 27,72 (18) |
| **Mortality, n (%)** | | |
| 30-day mortality | 749 (2) | 320 (2) |
| 90-day mortality | 1,350 (4) | 475 (3) |

Abbreviations. SD, standard deviation; IQR, interquartile range; AKI, acute kidney injury, ICU, intensive care unit.
[a] Baseline MAP was calculating as the median non-invasive MAP within the 48 hours prior to the first surgery start date-time that was between 60 and 110 mmHg.
[b] Values were calculated for 15-minute resampled series.

*Model Evaluation*

We trained the D3QN based RL models on a development cohort of 34,186 surgeries, resampled to 545,965 15-minute epochs. All results were reported from a test cohort of 15,835 surgeries, resampled to 255,748 15-minute epochs. We illustrated different aspects of agent suggestion evaluation in Figures 2-5. Action space distribution in Fig. 2 suggests that the distributions of the actions recommended by the agent and actions taken by the physicians have a high degree of similarity. The average Q-value distributions for surgeries that were grouped by the presence or absence of postoperative AKI were illustrated in Fig. 3 (A). We observed that surgeries with postoperative AKI tended to have lower return values, and in contrast, sessions without AKI concluded in higher return values. Fig. 3 (B) presents the relationship between physicians' actions and postoperative AKI within the first 3 days following surgery. In Fig 3 (B), physicians' treatments with a higher return value correspond to lower AKI probability, while treatments with a low return led to higher AKI probability.

On average, the RL model recommends less vasopressors and more IV fluids. The model replicated the physicians' vasopressor dosing decisions in 69% of cases, and it recommended a higher dose in 21% and a lower dose in 10% of treatments. For IV fluids, the actual dose was within 0.05 ml/kg per 15 min of the model's suggestion in 41% of cases; in the remainder, the model's recommendation was higher in 32% and lower in 27% of treatments.

Fig. 4 shows that administering either treatment at doses higher or lower than those recommended by the AI policy was associated with an increased probability of AKI. We presented the distribution of the estimated policy value of the physician's actual treatments, the RL policy, a random policy and a zero-drug policy in Fig. 5 (A), where estimated policy value is the expected cumulative reward each policy would yield over time. In this figure, the RL policy resulted in a higher estimated value compared to the alternatives. The order for the most influential variables in the RL model was given Fig. 5 (B). We identified Charlson comorbidity index and age as the top two important features in the decision-making process.

*Example Surgeries*

We illustrated four example surgical sessions including cases with and without postoperative AKI within the first three days after surgery. It is notable that these two example patients did not experience hypotensive episodes, and we observe the physician tended to administer IV fluids more and vasopressors almost each hour when MAP fluctuated around 65 mmHg compared to the case where the 15-minute average MAP remained consistently above 90 mmHg (Supplemental Figures 3 and 4). In both cases, the RL model did not recommend vasopressor administration. Similarly, compared to the RL model, the physician administered more IV fluids in the former case, while the RL model recommended more fluids in the latter. In the hypotensive example with postoperative AKI, the RL model recommends higher amounts of IV fluids from the start and maintains this level throughout most of the surgery compared to the physician's administration. Additionally, the RL model recommends the use of vasopressors, whereas the physician chose not to administer them. In the case of a non-hypotensive surgical patient who had postoperative AKI, the RL model's recommendations aligned with the physicians' no vasopressors administration decision along with similar IV fluid administration on average. (Supplemental Figures 5 and 6).

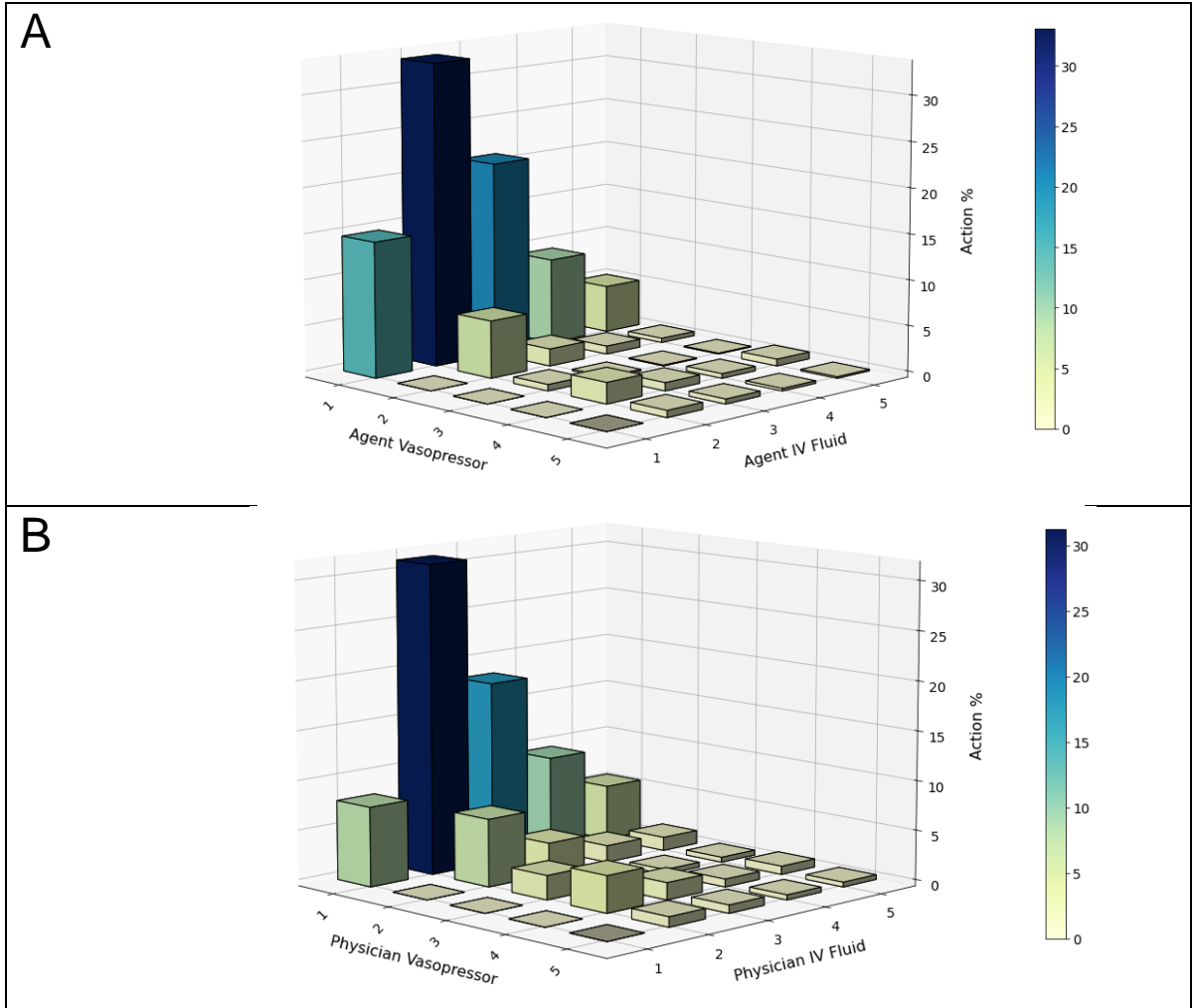

**Figure 2. Comparison of actions proposed by the trained agent (A) and taken by physician agent (B).** Each bin represents the tuple for discretized IV fluids and vasopressor actions.

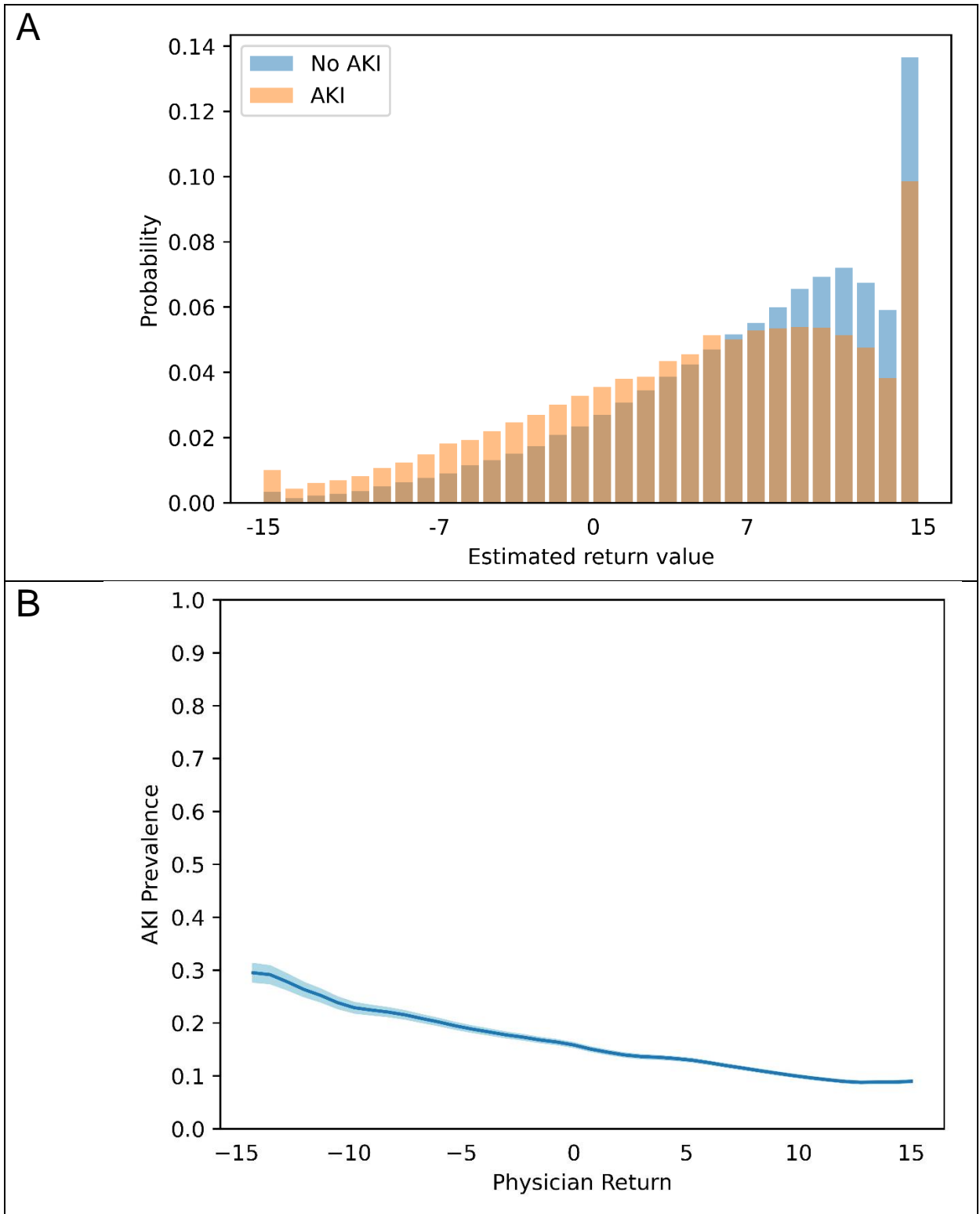

**Figure 3. Average return per surgery (A) and the relationship between the return of physicians' treatments and postoperative AKI within the first 3 days following surgery in test set (B).** (A) Distribution of average returns for cases with and without postoperative AKI. (B) Relationship between physician's return values and AKI prevalence in the test set.

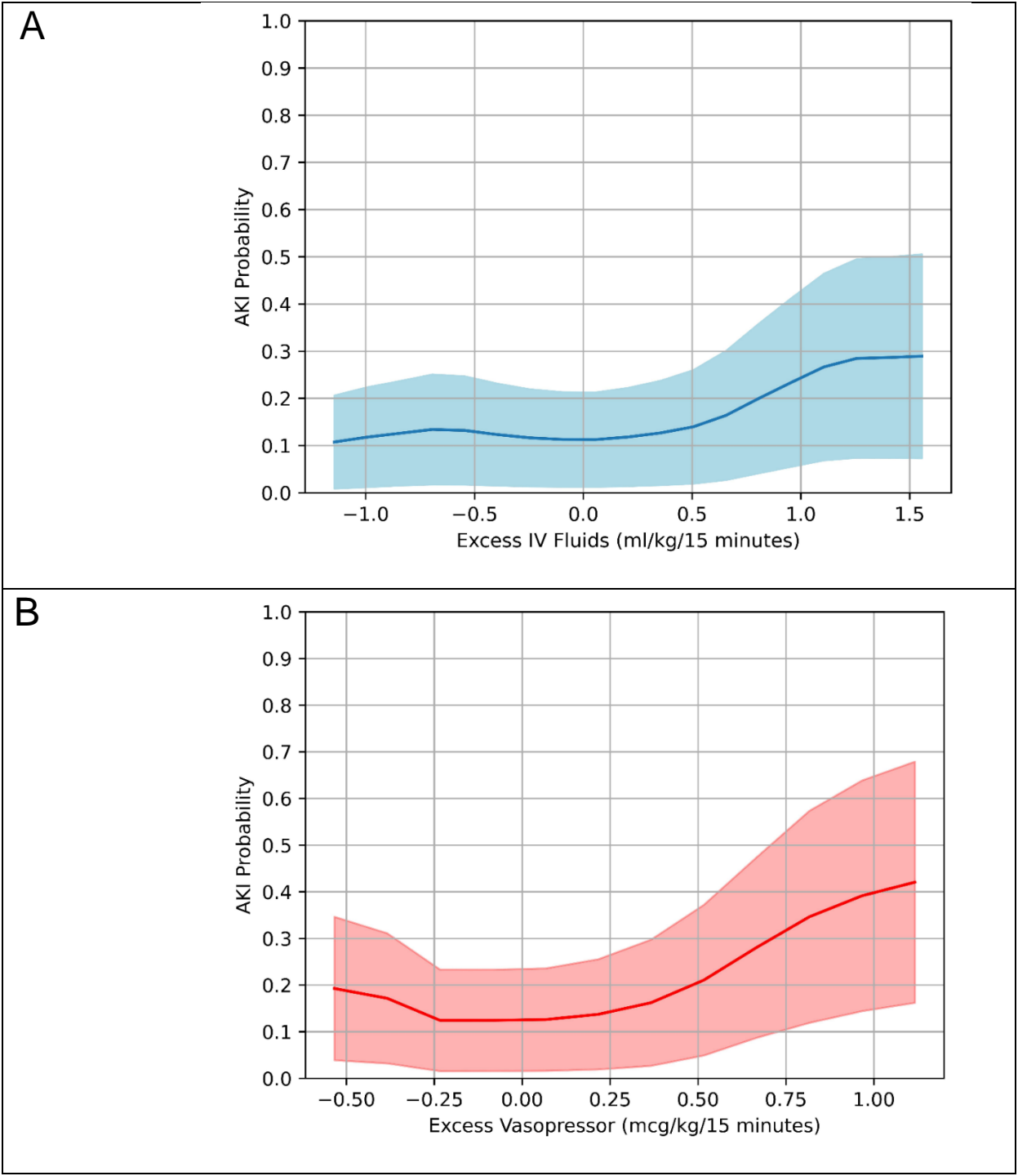

**Figure 4.** The average dose excess, calculated as the difference between given and recommended dose over all 15-minute intervals for IV fluids (A) and for vasopressors (B) per surgery.

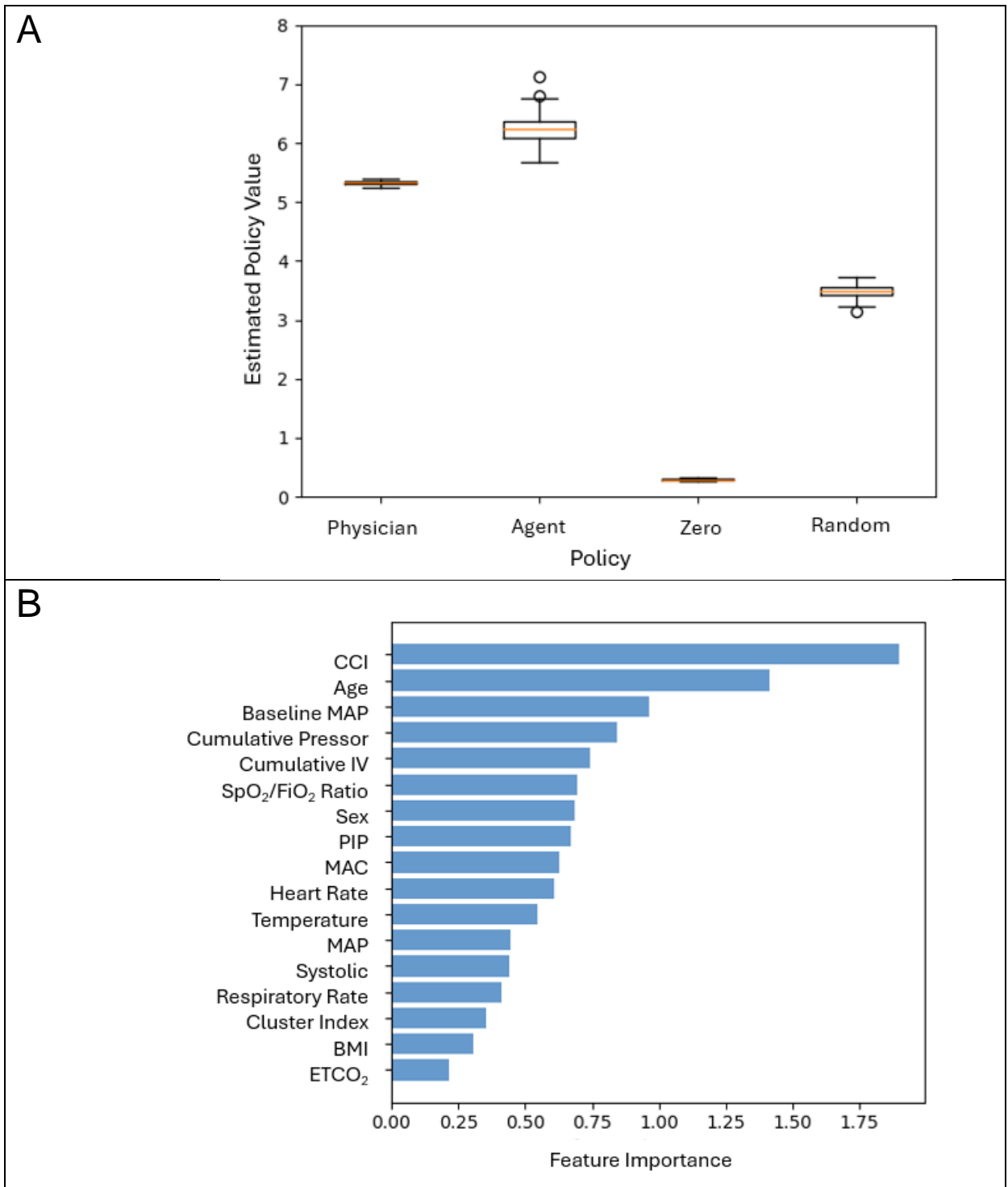

**Figure 5. Distribution of estimated values for physicians, agent, zero and random policies (A), feature importance derived for the trained model (B).** (A) Comparison of estimated policy value by WIS for physician policy, agent policy, zero policy, and random policy with 50 different K-Means initializations. (B) Feature importance obtained for RL model

**Discussion**

Postoperative AKI is prevalent and occurs in almost one in five patients following a major surgery.[7] Intraoperative hypotension affects approximately one in three patients undergoing non-cardiac surgical interventions.[42] Prior research has demonstrated strong associations between intraoperative hypotension and postoperative AKI.[43] Improving the management of intraoperative hypotension during non-cardiac surgery could save hospitals up to $4.6 million annually, primarily by decreasing the incidence of postoperative AKI.[44]

In this study, we introduced a RL modeling approach aimed to assist physicians in preventing intraoperative hypotension and reducing the risk of postoperative AKI by providing optimal IV fluid and vasopressor dosage recommendations. The model demonstrated promise in recommending actions that are associated with improved outcomes. While the actions recommended by the agent and the actions taken by physicians had a high degree of similarity, administering either IV fluids or vasopressors higher doses than those recommended by the AI policy was associated with an increased probability of postoperative AKI, and administering vasopressors at lower doses than recommended by the AI policy was also (albeit weakly) associated with an increased probability of postoperative AKI. We showed that the RL policy resulted in a higher estimated value compared to the alternatives, including treatments administered by physicians.

As previously presented, there have been studies dedicated to this issue, but their effectiveness relies heavily on the correctness of physician actions, which may replicate non-evidence-based practices in some cases, yielding development of an unstable policy. One possible solution could involve comparison of physician actions to actions suggested by multiple subject matter experts after thorough review of a sample of physician actions in various scenarios, especially scenarios in which patients experienced better or worse than expected outcomes. There exists no optimal, one-size-fits-all approach to clinical decision-making.[17,45]

Due to the complex, high-stakes, and often uncertain nature of surgical decisions for patients' with varying characteristics, a collaborative approach to shared decision-making involving the patient and all members of a clinical care team can improve patient satisfaction and may reduce costs associated with unnecessary treatments.

Although our results show promise, there are numerous challenges associated with implementing such a method in healthcare, especially in perioperative settings, unlike applications that can be tested in simulated environments where RL performance can be easily quantified. Unlike synthetic environments, clinical settings do not permit real-time experimentation with unvalidated policies. Thorough model validation is required to ensure patient safety. The literature describing the application of RL in healthcare focuses on its usage in dynamic environments to optimize treatment regimens. These dynamic environments, such as sepsis, mechanical ventilation, or glycemic decompensation, require that current models adapt rapidly to evolving patient states to avoid decompensation and physiological failure. Thus, one challenge to RL algorithms is the frequency of treatment recommendations. In their model for sepsis, another condition partially defined by hypotension, Komorowski et al.[20] and Tang et al.[30] aggregated patient data every four hours for sepsis treatment. The shortest timestep used in current literature was one hour.[21,24,31] Zhang et al.[46] utilized a model that specifically identified highly variable decision-making points amongst physicians to make treatment recommendations. Yu et al.[24] used 10-minute intervals to accomplish optimal mechanical ventilation, a treatment that is similar to fluid resuscitation in its timing and necessity. Peine et al.[47] used 4-hour time steps to dynamically optimize mechanical ventilation regime for critically ill patients. Prasad et al.[48] used a 10-minute timestep, but they optimize mechanical ventilation and sedation dosage and weaning from mechanical ventilation. In this work, we chose a time interval of 15 minutes under the assumption that intraoperative hypotension requires rapid correction, especially in a perioperative environment.

Future implementations of reinforcement learning in surgical settings should incorporate dynamic reward functions to accept input from both the patient and all members of a perioperative care team. Such collaborative reward functions can balance the risk aversion of individual patients and physicians with the expected benefits and postoperative care trajectories highlighted by other team members. By giving the patient increased control over their own algorithm-influenced clinical care, collaborative and dynamic reward functions can lead to an increase in overall patient satisfaction.

**Conclusion**

We present a reinforcement learning modeling approach to recommend optimal intravenous fluid and vasopressor doses to avoid intraoperative hypotension and postoperative acute kidney injury. When clinician actions most closely imitated model recommendations, AKI incidence was lowest. These findings require prospective validation and clinical implementation to assess the potential to improve perioperative care and patient outcomes.

**Acknowledgements**

We acknowledge the University of Florida Integrated Data Repository (IDR) and the University of Florida Health Office of the Chief Data Officer for providing the analytic data set for this project. Additionally, the Research reported in this publication was supported by the National Center for Advancing Translational Sciences of the National Institutes of Health under University of Florida Clinical and Translational Science Awards UL1 TR000064 and UL1TR001427.

**Funding**

T.O.B. was supported by K01 DK120784 from the National Institute of Diabetes and Digestive and Kidney Diseases (NIH/NIDDK). TOB received grant (97071) from Clinical and Translational Science Institute, University of Florida and Research Opportunity Seed Fund grant (DRPD-ROSF2023) from University of Florida Research. Additionally, the Research reported in this publication was supported by the National Center for Advancing Translational Sciences of the National Institutes of Health under University of Florida Clinical and Translational Science Awards UL1 TR000064 and UL1TR001427. The Titan X Pascal partially used for this research was donated by the NVIDIA Corporation. The content is solely the responsibility of the authors and does not necessarily represent the official views of the National Institutes of Health. The funders had no role in study design, data collection and analysis, decision to publish, or preparation of the manuscript. AB and TOB have full access to all the data in the study and take responsibility for the integrity of the data and the accuracy of the data analysis.

**Author contributions**

AB and TOB have full access to the data in the study and take responsibility for the integrity of the data and the accuracy of the data analysis. TOB, PR, BS, TJL, TL, DN, EA, and YR contributed to the study design. EA, TL, DN, HL, TJL, and TOB drafted the manuscript. EA, TL, DN, HL, and MR worked on data processing. Analysis was performed by EA, TL, and DN.

Funding was obtained by TOB and TJL. Administrative, technical, and material support was provided by AB and PR. All authors contributed to the interpretation of data and to critical revision of the manuscript for important intellectual content.

## Additional information

### Competing interests

# Supplemental Materials

**Learning optimal treatment strategies for intraoperative hypotension using deep reinforcement learning**

Esra Adiyeke[*], Tianqi Liu[*], Venkata Sai Dheeraj Naganaboina[*], Han Li, Tyler J. Loftus, Yuanfang Ren, Benjamin Shickel, Matthew M. Ruppert, Karandeep Singh, Ruogu Fang, Parisa Rashidi, Azra Bihorac[#], Tezcan Ozrazgat-Baslanti[#]

* These authors have contributed equally as first authors

# These authors have contributed equally as senior authors

This supplemental material has been provided by the authors to give readers additional information about their work.

**Supplemental Material Table of Contents**

**Supplemental Methods**

# Supplementary Methods

## A. Cohort
Data from the full patient cohort were divided into a development cohort, comprising patients admitted between June 1, 2014 and November 29, 2018, and a validation cohort, from November 30, 2018 to September 20, 2020 in a 70% to 30% ratio. Patients were excluded if they had any of the following cases: 1) end stage kidney disease present, 2) age < 18 on admission, 3) not an inpatient encounter, 4) missing admission or discharge date or missing surgery stop time, 4) surgery performed for organ donation, a minor gastrointestinal procedure, or a minor pain management procedure, anesthesia administered outside of the operating room, 5) the surgery was < 60 minutes, 6) 3-day or 7-day acute kidney injury (AKI) status was missing due to insufficient serum creatinine data available, 7) cardiac surgeries. (Supplemental Figure 1).

## B. Dosage pre-processing and action space
Intravenous fluids (IV) included boluses and continuous infusions of crystalloid and colloid solutions. Vasopressors included dopamine, epinephrine, norepinephrine, phenylephrine, and vasopressin. Outliers were capped to expert defined clinically plausible values identified utilizing maximum doses used in refractory shock (Supplemental Table 3). Vasopressor dosages were converted into norepinephrine-equivalents using dosage conversions as listed in Supplemental Table 3.

Using norepinephrine-equivalent dosages normalized to rate and weight, we developed a discrete 5 x 5 action space of vasopressors and IV fluid dosages given in 15-minute intervals. Thresholds for action bins were chosen by selecting percentile thresholds for points of high rate of dosage increase and were validated through expert review. We converted each drug dosage at every timestep into an integer representing its bin. No drug dosage was encoded as bin 1, the lowest dosage category was encoded as bin 2 and the highest dosage category was encoded as bin 5. Thus, interventions were represented as tuples of total IV and vasopressor dosages per 15-minute interval.

## C. Model Details
Surgery can be conceptualized as a sequential decision-making process, and to train an agent, we used the Reinforcement Learning Algorithm with 16 clinical variables (age, sex, Charlson comorbidities index, body mass index, heart rate, systolic blood pressure, mean arterial pressure, body temperature, respiratory rate, minimum alveolar concentration, SF ratio, peak inspiratory pressure, end-tidal carbon dioxide, average mean arterial blood pressure of past 48 hours, cumulative IV and cumulative pressor administered during surgery). The physiological state of a patient was represented by these 16 variables, and thus the surgical decision-making process could be formulated as a partially observable Markov Decision Process (MDP). The model free MDP was defined by a tuple $\{S, A, R, \gamma\}$ with:
1. $S$ is the state of a patient (in our model, it contains 16 clinical variables)
2. $A$ is the finite set of actions for state $S$ (in our model, the doses intravenous fluids and vasopressors are discretized into 25 actions)

3. $R(s')$ is the immediate reward received for transitioning to next state $S'$.
4. $\Gamma$ is the discounting factor, that indicates the decay of influences for future rewards than an immediate reward.

Our model operated on 15-minute time intervals, within which multiple measurements were recorded. We considered resampling by averaging all measurements within each interval as a preprocessing step before inputting the data into the model. We set $\Gamma$ to 0.99 in the model.

We utilized the K-means++ clustering algorithm to categorize the resampled data points into distinct clusters, each representing a unique patient state. Specifically, we used a total of 16 features derived from the patient data for clustering, to provide a comprehensive representation of the patient's condition. The K-means++ algorithm was chosen for its efficiency in selecting initial cluster centroids, minimizing the chances of poor clustering due to suboptimal centroid initialization. Cluster membership was assigned based on the closest centroids, ensuring that each data point was grouped with the most similar patient states. We determined an optimal number of 200 clusters using the elbow method for silhouette analysis, which effectively discretized our state space into a manageable set of 200 potential states. We developed 50 distinct RL models for 50 different clustering outcomes obtained by different random initializations to account for the variability in policy values. This discretization allows us to group patients with similar medical conditions and tailor specific treatment strategies for each group. Additionally, the cluster number was included as a feature in the overall feature set before training, providing our model with an additional layer of insight into patient state transitions. We constructed the transition matrix $T(s',s,a)$ by analyzing the frequency of observed transitions in the training dataset, then normalized these frequencies to obtain proper transition probabilities.

**Reward Function**

We developed a reward function as a combination of long- and short-term rewards with two major parts to consider. The reward function is the sum of rewards of two parts:

$$r = r_{aki} + r_{hypo} \qquad \text{Eq. 1}$$

Part 1: Postoperative AKI outcome: This acts as the long-term reward/penalty utilizing a dedicated reward of +15 if no AKI occurred within 3 days following surgery, and -15 for if AKI occurred within 3 days following surgery.

$$r_{aki} = \begin{cases} 15, & \text{no AKI within 3 days of surgery} \\ -15, & \text{AKI within 3 days of surgery} \end{cases} \qquad \text{Eq. 2}$$

Part 2: Hypotension: The function adds a penalty of 1.75 ($c_1$) if the state is hypotensive, that is mean arterial pressure (MAP) is less than 65 mmHg or 20% lower than baseline MAP. The baseline MAP was calculated by taking the median non-invasive MAP within the 48 hours prior to the first surgery start date-time that was between 60 and 110 mmHg.

$$r_{hypo} = \begin{cases} -1.75, & MAP_t < 65 mmHg \text{ or } MAP_t < 0.8 \times MAP_{baseline} \\ 0, & otherwise \end{cases} \qquad \text{Eq. 3}$$

**Evaluation of physicians' action:**

We used Dueling Double Deep Q Networks (D3QN) as architecture for the RL model to learn an optimal policy (agent policy) and test it. We performed an evaluation of the real actions (the policy) of physicians using temporal difference (TD) for the Q function learning with observed states, actions and rewards tuples in surgeries. The Q function learning is computed iteratively by the following formula:

$$Q^\pi(s,a) \leftarrow Q^\pi(s,a) + \alpha\,(r + \gamma Q^\pi(s',a') - Q^\pi(s,a))\ ,\qquad \text{Eq. 4}$$

where the $\alpha$ is the learning rate and $r$ is the immediate reward.

The loss function of the D3QN is denoted as:

$$L_{DQN} = \left(\left(r + \gamma \max_{a_{t+1}} Q(s_{t+1}, a_{t+1},;\theta^{target})\right) - Q(s,a;\theta^{pred})\right)^2 \qquad \text{Eq. 5}$$

The training experiences are recorded from physicians, therefore the action recommended by AI agent should have similar distribution with actions from physicians to ensure the safety of the agent policy. We added a KL divergence penalty to the loss function to constrain the action distribution for AI agent from diverging from physicians' actions:

$$L = L_{DQN} + \alpha L_{KL} \qquad \text{Eq. 6}$$

$$L_{KL} = D_{KL}(D_{Physician}||D_{Agent}) = \sum_{i=1}^{N}[p_{phy}(x_i)\log p_{phy}(x_i) - p_{phy}(x_i)\log q_{agent}(x_i)] \qquad \text{Eq. 7}$$

**Estimation of the AI policy**

Our model learned an optimal policy, as we refer as agent policy, in theory with the goal of maximizing the sum of rewards to avoid long-term AKI and short-term Hypotension. The agent policy will start with a random policy that was iteratively evaluated and then improved until converging to an optimal solution. After convergence, the agent policy $\pi^*$ corresponded to actions with the highest state-action value:

$$\pi^*(s) \leftarrow argmax_a Q^{\pi^*}(s,a),\ \forall s \qquad \text{Eq. 8}$$

**D. Model Evaluation**

**Weighted Importance Sampling (WIS)**

The traditional reinforcement learning applications have typically been tested in simulated environments where the evaluation of performance of reinforcement learning algorithms can be easily quantified, such as in video games. However, unlike synthetic environments, clinical settings do not permit real-time experimentation with unvalidated policies. Off-policy evaluation allows for retrospective assessment of a target policy (e.g., model suggested treatment strategy) under a different behavior policy (e.g., physician decision making). Consequently, off-policy evaluation is adopted for our task to evaluate our trained AI agent ($\pi_e$) based on physicians' policies. Weighted Importance Sampling (WIS) takes the idea to reweight the rewards in the historical data (physicians' policy $\pi_b$) by the importance sampling ratio between $\pi_e$ and $\pi_b$.

The per-step importance ratio is defined at step *t* as:

$$\rho_t = \frac{\pi_e(a_t|s_t)}{\pi_b(a_t|s_t)} \qquad \text{Eq. 9}$$

The cumulative importance ratio up to step t is formulated as:

$$\rho_{1:t} = \prod_{t'=1}^{t} \rho_{t'}\ , \qquad \text{Eq. 10}$$

where $t$ is the end of a surgery. However, the cumulative importance ratio is not stable for long surgeries due to the large number of steps; that is 15-min intervals. When most $\rho_t$ are larger than 1, the product value will soon exponentially explode to a very large number, which makes the evaluation not stable. To solve this problem, we proposed a new cumulative ratio as follows:

$$\rho'_{1:t} = ln\rho_{1:t} = ln \prod_{t'=1}^{t} \rho_{t'} \qquad \text{Eq. 11}$$

The $ln$ can compress the range of input value. But if the $\rho_{1:t}$ is smaller than 1, the output of $ln$ function is smaller than 0. Since a negative cumulative ratio is not meaningful in this context, we instead used the Softmax function, which maps values from the range $(-\infty, \infty)$ to $[0, 1]$.

$$\begin{aligned}
\rho''_{1:t} &= \frac{1}{1+e^{-\rho'_{1:t}}} \\
&= \frac{1}{1+e^{-ln\rho_{1:t}}} \\
&= \frac{1}{1+\frac{1}{\rho_{1:t}}}
\end{aligned} \qquad \text{Eq. 12}$$

We then calculated the average cumulative importance ratio for a dataset $D$ with $|D|$ number of trajectories, that is the number of surgeries in this case:

$$w_D = \frac{\sum_{i=1}^{|D|} \rho''^{(i)}_{1:t_i}}{|D|} \qquad \text{Eq. 13}$$

The trajectory-wise WIS estimator is given by:

$$V^{(i)}_{WIS} = \frac{\rho''^{(i)}_{1:t^i}}{w_D} \left( \sum_{t=1}^{t^i} \gamma^{t-1} r_t \right) \qquad \text{Eq. 14}$$

Then the WIS estimator was calculated as the average estimate over all trajectories:

$$\text{WIS} = \frac{1}{|D|} \sum_{i=1}^{|D|} V^{(i)}_{WIS} \quad , \qquad \text{Eq. 15}$$

where $V^{(i)}_{WIS}$ is WIS applied to the i-th trajectory.

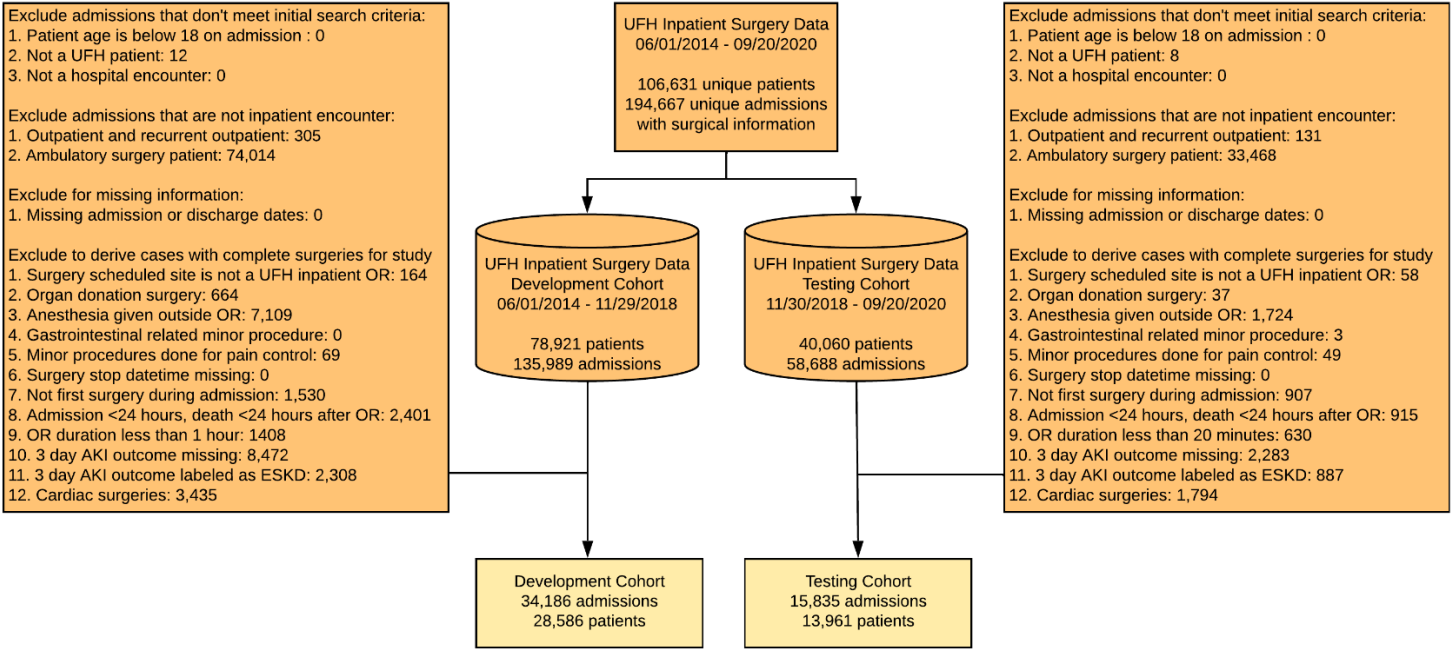

**Supplemental Figure 1.** Figure illustrating derivation of the study population

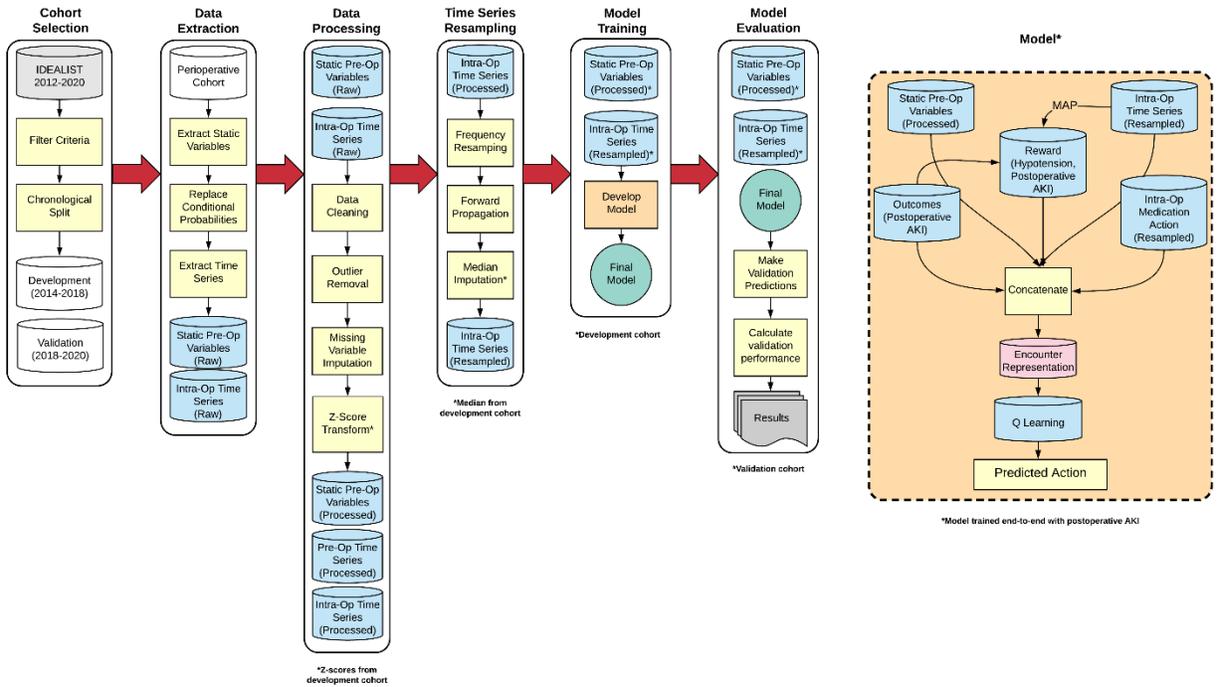

**Supplemental Figure 2.** Workflow pipeline (Abbreviations: AKI, acute kidney injury)

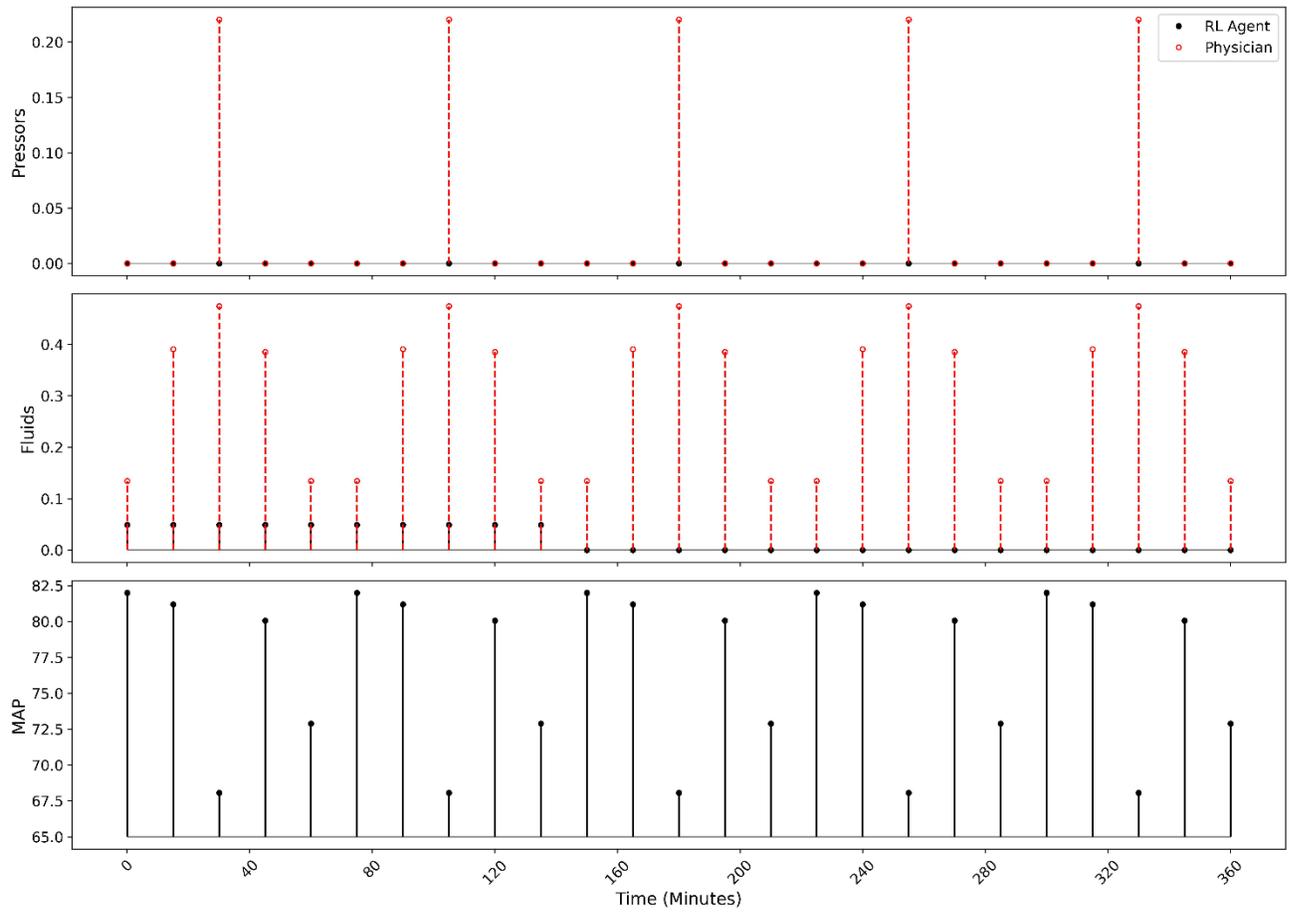

**Supplemental Figure 3.** Comparison between doses suggested by RL model and physician's administration for a surgical session without postoperative AKI within the first 3 days after surgery. (Example surgery 1)

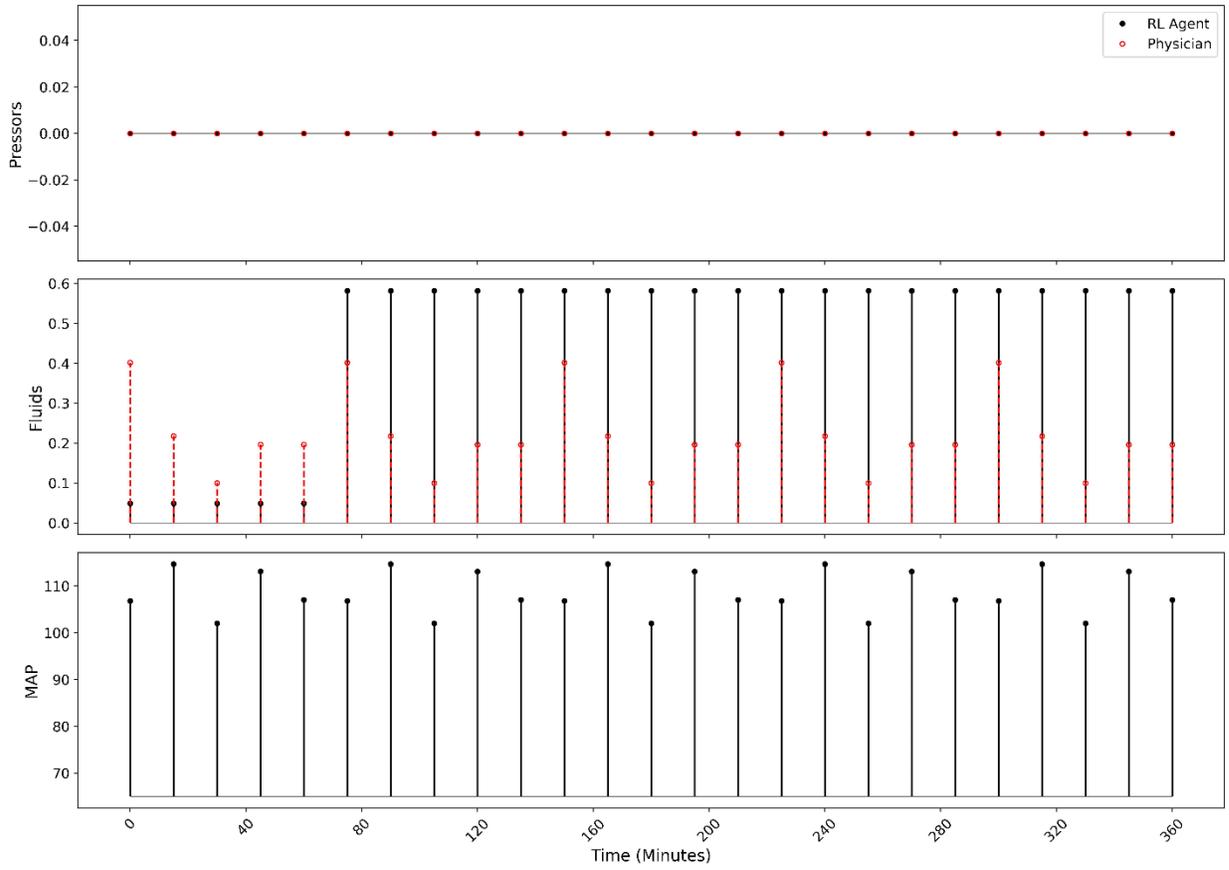

**Supplemental Figure 4.** Comparison between doses suggested by RL model and physician's administration for a surgical session without postoperative AKI within the first 3 days after surgery. (Example surgery 2)

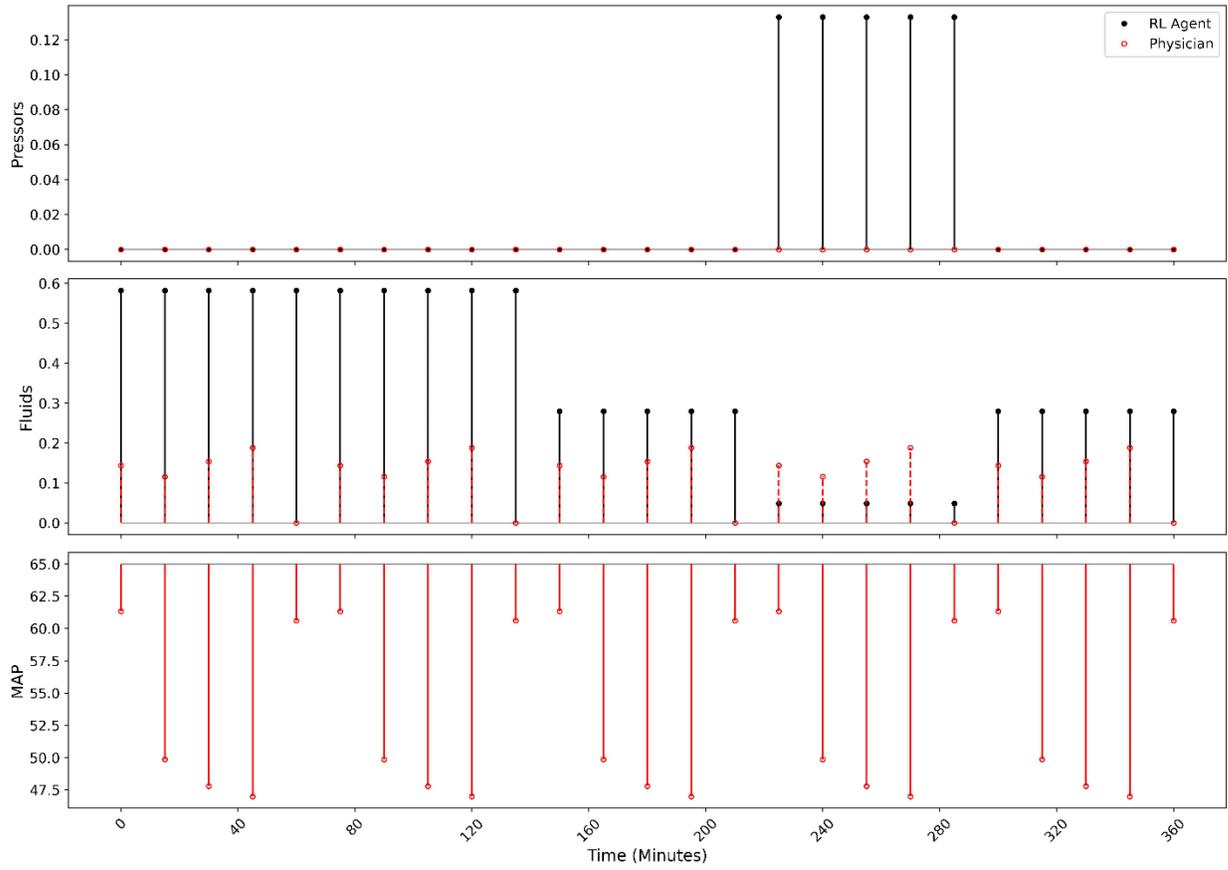

**Supplemental Figure 5.** Comparison between doses suggested by RL model and physician's administration for a surgical session with postoperative AKI within the first 3 days after surgery. (Example surgery 3)

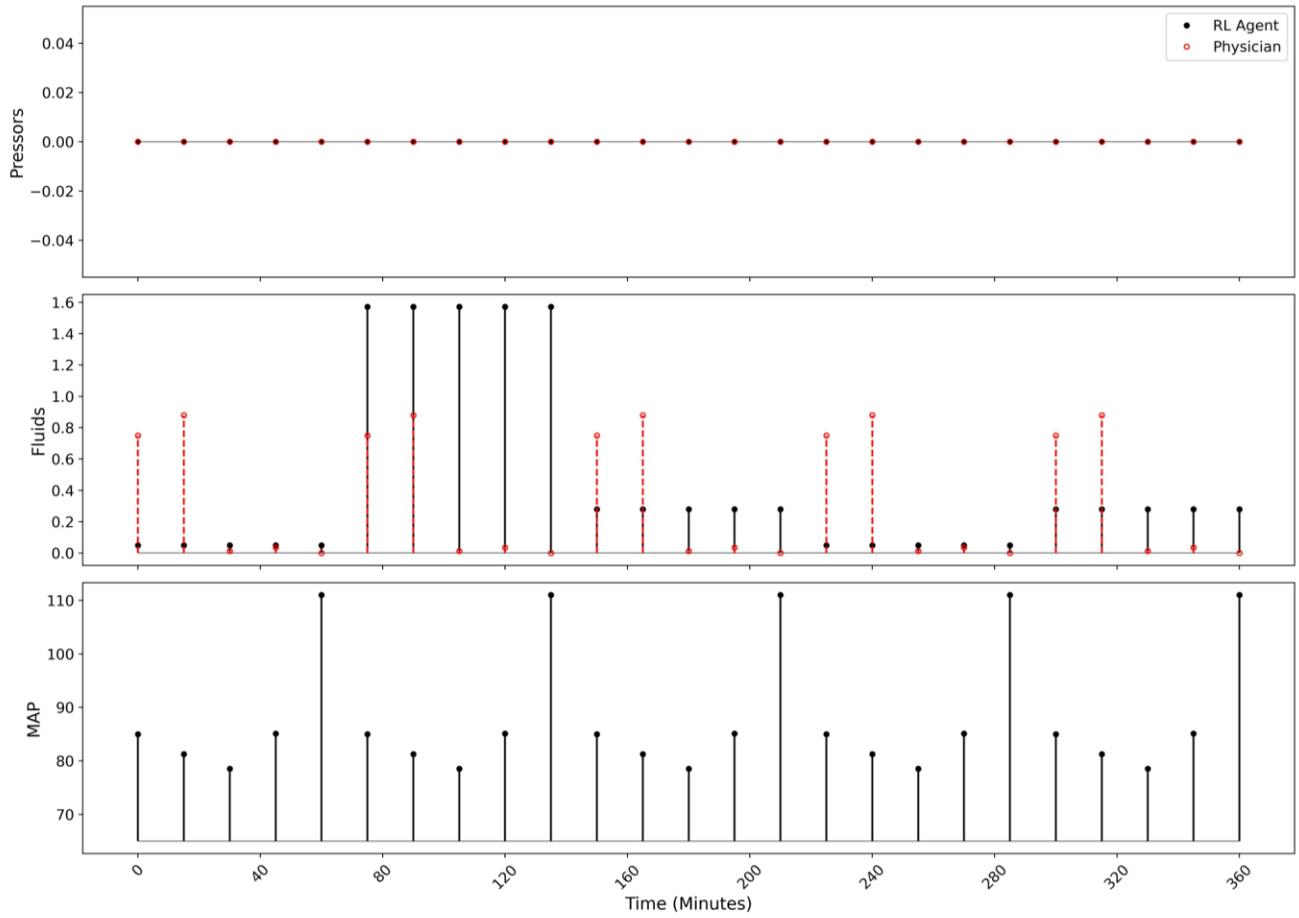

**Supplemental Figure 6.** Comparison between doses suggested by RL model and physician's administration for a surgical session with postoperative AKI within the first 3 days after surgery. (Example surgery 4)

**Supplemental Table 1:** Table listing characteristics of input variables.

| Variable | Type of Variable | Data Source | Number of Categories | Type of Preprocessing |
|---|---|---|---|---|
| **Demographic Variables** | | | | |
| Age (years) | Continuous | Derived | | Imputation of outliers[a] |
| Sex | Binary | Raw | 2 | |
| Body Mass Index | Continuous | Raw | | Imputation of outliers[a] |
| **Comorbidities** | | | | |
| Charlson's Comorbidity Index | Nominal | Derived | 18 | Optimization of categorical features[b] |
| **Intraoperative Medications[c]** | | | | |
| Vasopressors | Continuous | Raw | | Data cleaning[d] |
| Intravenous (IV) fluid | Continuous | Raw | | Data cleaning[d] |
| **Physiological Intraoperative Time Series** | | | | |
| Systolic blood pressure, mmHg | Continuous | Raw | | Data cleaning[d] |
| Mean Arterial Pressure, mmHg | Continuous | Raw | | Data cleaning[d] |
| Minimum alveolar concentration | Continuous | Raw | | Data cleaning[d] |
| Heart rate, bpm | Continuous | Raw | | Data cleaning[d] |
| Temperature (°C) | Continuous | Raw | | Data cleaning[d] |
| End-tidal CO2 (ETCO2) | Continuous | Raw | | Data cleaning[d] |
| Peak Inspiratory Pressure (PIP) | Continuous | Raw | | Data cleaning[d] |
| Respiratory Rate | Continuous | Raw | | Data cleaning[d] |
| SPO2/FIO2 ratio | Continuous | Derived | | Data cleaning[d] |

A different set of variables was kept in final models (preoperative or intraoperative) from the input set provided in the table.

[a] For continuous variables, observations that fell in the top and bottom 1% of the distribution were considered as outliers and imputed by neighborhood values (i.e., above 99%) are imputed randomly from a uniform distribution defined over [95%, 99.5%] percentiles and below 1% are imputed randomly from another uniform distribution defined over [0.5%, 5%] percentiles.

[c] Medications were taken during surgery using RxNorms data grouped into drug classes according to the US, Department of Veterans Affairs National Drug File-Reference Terminology.

[d] We used observations for the first surgery, in case multiple surgeries exist. We averaged values if multiple observations exist at a time point.

**Supplemental Table 2:** Detailed cohort characteristics.

| Features | Development Cohort | Test Cohort |
|---|---|---|
| Number of encounters, n | 34,186 | 15,835 |
| **Demographic information** | | |
| Age, years, mean (SD) | 57 (17) | 59 (17) |
| Sex, n (%) | | |
|   Male, n (%) | 17,031 (50) | 7,909 (50) |
|   Female, n (%) | 17,155 (50) | 7,926 (50) |
| Race, n (%) | | |
|   White | 26,743 (78) | 12,360 (78) |
|   African American | 4,776 (14) | 2,233 (14) |
|   Other | 2,164 (6) | 972 (6) |
|   Missing | 503 (1) | 270 (2) |
| Ethnicity, n (%) | | |
|   Non-Hispanic | 32,063 (94) | 14,711 (93) |
|   Hispanic | 1530 (4) | 756 (5) |
| Body Mass Index, median (IQR) | 28 (24, 34) | 28 (24, 34) |
| **Comorbidities, n (%)** | | |
|   Charlson comorbidity index, median (IQR) | 4 (2, 6) | 4 (2, 6) |
|   Alcohol or drug abuse | 5,353 (16) | 2,587 (16) |
|   Myocardial Infarction | 2,384 (7) | 1,156 (7) |
|   Congestive Heart Failure | 4,653 (14) | 2,488 (16) |
|   Peripheral Vascular Disease | 6,967 (20) | 3,644 (23) |
|   Cerebrovascular Disease | 5,642 (17) | 2,854 (18) |
|   Chronic Pulmonary Disease | 1,0516 (31) | 5,333 (34) |
|   Cancer | 1,0186 (30) | 4,465 (28) |
|   Metastatic Carcinoma | 3,451 (10) | 1,678 (11) |
|   Liver Disease | 5,232 (15) | 2,549 (16) |
|   Diabetes | 8,252 (24) | 3,916 (25) |
|   Hypertension | 21,843 (64) | 10,642 (67) |
|   Obesity | 11,393 (33) | 7,070 (45) |
|   Fluid and electrolyte disorders | 10,301 (30) | 6,120 (39) |
|   Valvular Disease | 3,561 (10) | 2,048 (13) |
|   Coagulopathy | 4,674 (14) | 2,173 (14) |
|   Weight Loss | 5,212 (15) | 2,589 (16) |
|   Depression | 9,544 (28) | 4,805 (30) |
|   Chronic Anemia | 6,792 (20) | 4,183 (26) |
|   Chronic Kidney Disease | 5,718 (17) | 3,019 (19) |
| Reference estimated glomerular filtration rate, median (IQR) | 95.27 (81.86, 109.73) | 93.44 (80.53, 107.71) |

| Features | Development Cohort | Test Cohort |
|---|---|---|
| Baseline mean arterial pressure mmHg, median (IQR)[a] | 86 (78, 94) | 87 (79, 95) |
| **Surgical information, n (%)** | | |
| Time from Admission to Surgery, days, median (IQR) | 3 (2, 22) | 3 (2, 23) |
| Emergency admission | 12,243 (36) | 5,763 (36) |
| Admission type | | |
| Surgery | 13,080 (38) | 6,521 (41) |
| Medicine | 16,796 (49) | 6,920 (44) |
| Transferred from another hospital | 5,666 (17) | 2,667 (17) |
| Anesthesia Type | | |
| General | 31,281 (92) | 14,941 (94) |
| Local/regional | 2,905 (8) | 894 (6) |
| Surgery Type | | |
| Orthopedic Surgery | 10,109 (30) | 4,341 (27) |
| Neurosurgery | 3,744 (11) | 2,546 (16) |
| Vascular Surgery | 3,420 (10) | 1,704 (11) |
| Other | 3,234 (9) | 2,029 (13) |
| Urology | 3,768 (11) | 1,231 (8) |
| Gastrointestinal Surgery | 2,798 (8) | 1,165 (7) |
| Ear Nose Throat | 1,873 (5) | 855 (5) |
| OB Gynecology | 1,291 (4) | 541 (3) |
| Surgical Oncology | 1,495 (4) | 449 (3) |
| Plastic Surgery | 619 (2) | 277 (2) |
| Transplantation | 575 (2) | 170 (1) |
| Burn Surgery | 883 (3) | 321 (2) |
| Pediatric Surgery | 270 (1) | 122 (1) |
| Ophthalmology | 95 (0) | 79 (0) |
| Medicine Gastroenterology | 12 (0) | 5 (0) |
| **Intraoperative vitals** | | |
| Systolic blood pressure | | |
| Measured value, mmHg, median IQR | 114.0 (102.0, 130.0) | 116.0 (104.0, 132.0) |
| Total measurements, n, IQR | 73.0 (47.0, 149.0) | 81.0 (50.0, 179.0) |
| Encounters missing, n (%) | 21 (0.06) | 11 (0.07) |
| Mean arterial pressure | | |
| Measured value, mmHg, median IQR | 79.0 (70.0, 90.0) | 82.0 (72.0, 93.0) |
| Total measurements, n, IQR | 73.0 (47.0, 150.0) | 81.0 (50.0, 179.0) |
| Encounters missing, n (%) | 21 (0.06) | 11 (0.07) |
| Heart rate | | |
| Measured value, bpm, median IQR | 75.0 (65.50, 86.50) | 75.0 (66.0, 86.50) |
| Total measurements, n, IQR | 175.0 (121.75, 257.0) | 179.0 (124.0, 259.0) |
| Encounters missing, n (%) | 22 (0.06) | 9 (0.06) |
| Oxygen saturation (SpO2) | | |
| Measured value, %, median IQR | 99.10 (97.50, 10.0) | 99.0 (97.20, 10.0) |
| Total measurements, n, IQR | 177.0 (120.0, 268.0) | 184.0 (123.0, 277.0) |
| Encounters missing, n (%) | 43 (0.13) | 18 (0.11) |
| Fraction of inspired oxygen (FiO2) | | |
| Measured value, %, median IQR | 40.0 (40.0, 40.0) | 40.0 (40.0, 40.0) |

| Features | Development Cohort | Test Cohort |
|---|---:|---:|
| Total measurements, n, IQR | 187.0 (130.0, 280.0) | 195.0 (135.0, 291.0) |
| Encounters missing, n (%) | 19 (0.06) | 8 (0.05) |
| End-tidal carbon dioxide (EtCO2) | | |
| Measured value, mmHg, median IQR | 34.0 (32.0, 37.0) | 35.0 (33.0, 38.0) |
| Total measurements, n, IQR | 149.0 (82.0, 238.0) | 2.0 (0.0, 163.0) |
| Encounters missing, n (%) | 3,548 (10.38) | 7,254 (45.81) |
| Respiration rate | | |
| Measured value, breaths/minute, median IQR | 10.0 (8.0, 12.0) | 12.0 (10.0, 14.0) |
| Total measurements, n, IQR | 183.0 (125.0, 275.0) | 124.0 (3.0, 214.0) |
| Encounters missing, n (%) | 75 (0.22) | 782 (4.94) |
| Peak inspiratory pressure | | |
| Measured value, mmHg, median IQR | 18.0 (14.0, 23.0) | 18.0 (14.0, 22.0) |
| Total measurements, n, IQR | 183.0 (126.0, 276.0) | 176.0 (109.0, 271.0) |
| Encounters missing, n (%) | 396 (1.16) | 1,567 (9.90) |
| Minimum alveolar concentration | | |
| Measured value, median IQR | 0.62 (0.44, 0.81) | 0.56 (0.31, 0.77) |
| Total measurements, n, IQR | 160.0 (102.0, 242.0) | 173.0 (118.0, 252.0) |
| Encounters missing, n (%) | 3,234 (9.46) | 1,449 (9.15) |
| Core temperature | | |
| Measured value, degrees Celsius, median IQR | 36.83 (36.28, 37.33) | 36.94 (36.33, 37.44) |
| Total measurements, n, IQR | 126.0 (65.0, 202.0) | 106.0 (13.0, 184.0) |
| Encounters missing, n (%) | 1,811 (5.30) | 1,233 (7.79) |
| **Intraoperative medications, median (IQR)[b]** | | |
| Vasopressor total dose per 15 min (mcg/kg) | 0 (0, 0.04) | 0 (0, 0.08) |
| Intravenous fluids total dose per 15 min (ml/kg) | 0.12 (0.03, 0.39) | 0.13 (0.03, 0.37) |
| **Intraoperative hypotension, n (%)** | 24,144 (70) | 9,677 (61) |
| **Complications, n (%)** | | |
| Acute kidney injury | 5,649 (17) | 27,72 (18) |
| Acute kidney injury within 3 days after surgery | 3,870 (11) | 1,856 (12) |
| Acute kidney injury within 7 days after surgery | 4,637 (14) | 2,242 (14) |
| Cardiovascular complication | 4,509 (13) | 2,508 (16) |
| Neurological complication and delirium | 4,167 (12) | 2,615 (17) |
| Prolonged ICU stay | 8,533 (25) | 4,694 (30) |
| Prolonged mechanical ventilation | 2,155 (6) | 941 (6) |
| Sepsis | 2,891 (8) | 1,491 (9) |
| Venous thromboembolism | 1,592 (5) | 858 (5) |
| Wound complication | 5,389 (16) | 3,540 (22) |
| 30-day mortality | 749 (2) | 320 (2) |
| 90-day mortality | 1,350 (4) | 475 (3) |

Abbreviations. SD, standard deviation; IQR, interquartile range; ICU, Intensive Care Unit.

[a] Baseline MAP was calculating as the median non-invasive MAP within the 48 hours prior to the first surgery start date-time that was between 60 and 110 mmHg.
[b] Values were calculated for 15-minute resampled series.

**Supplemental Table 3.** Maximum allowed dose for outlier cleaning and conversion factors

| Vasopressor | Maximum Allowed Dose | Norepinephrine Equivalent Dose* |
|---|---|---|
| Norepinephrine | 3.3 mcg/kg/min | 1 |
| Epinephrine | 2 mcg/kg/min | 1 |
| Vasopressin | 0.1 units/min | 2.5 |
| Phenylephrine | 10 mcg/kg/min | 0.1 |
| Dopamine | 50 mcg/kg/min | 0.01 |

*Approximate conversion of vasopressin dose in units/min to equivalent norepinephrine dose in mcg/kg/min, normalized to 100 kg body weight.

Norepinephrine equivalent dose (mcg/kg/min) = dose (mcg/kg/min)*coefficient, for epinephrine, phenylephrine, and dopamine

Norepinephrine equivalent dose (mcg/kg/min) = dose (units/min)*coefficient*(weight in kg/100kg) for vasopressin

**Supplemental Table 4.** Intravenous (IV) fluid dose discretization

| IV fluid Action | 1 | 2 | 3 | 4 | 5 |
|---|---|---|---|---|---|
| Range of dose | 0 (None) | 0<x<=0.16 ml/kg/15 mins (0-50%) | 0.16<x<=0.43 ml/kg/15 mins (50-75%) | 0.43 < x<=0.90 ml/kg/15 mins (75-90%) | x>0.90 ml/kg/15 mins (90-100%) |

**Supplemental Table 5.** Vasopressor dose discretization

| Pressor Action | 1 | 2 | 3 | 4 | 5 |
|---|---|---|---|---|---|
| Range of dose | 0 (None) | 0<x<=0.29 mcg/kg/15 mins (0-50%) | 0.29<x<=0.51 mcg/kg/15 mins (50-75%) | 0.51 < x<=0.89 mcg/kg/15 mins (75-90%) | x>0.89 mcg/kg/15 mins (90-100%) |